\documentclass[12pt]{iopart}

\usepackage{epsfig}
\usepackage{color}
\usepackage[latin1]{inputenc}
\usepackage{float}
\usepackage{graphicx}
\newcommand{\be}{\begin{eqnarray}}
\newcommand{\ee}{\end{eqnarray}}

\newcommand{\benum}{\begin{enumerate}}
\newcommand{\eenum}{\end{enumerate}}

\usepackage{iopams}  
\begin{document}

\title{Flow in heavy-ion collisions - Theory Perspective}

\author{Bj\"orn Schenke}
\address{Physics Department, Building 510A, Brookhaven National Laboratory\\ Upton, NY 11973, USA}

\begin{abstract}
I review recent developments in the field of relativistic hydrodynamics and its application to the bulk dynamics in heavy-ion collisions at the Relativistic Heavy-Ion Collider (RHIC) and the Large Hadron Collider (LHC). In particular, I report on progress in going beyond second order relativistic viscous hydrodynamics for conformal fluids, including temperature dependent shear viscosity to entropy density ratios, as well as coupling hydrodynamic calculations to microscopic hadronic rescattering models. I describe event-by-event hydrodynamic simulations and their ability to compute higher harmonic flow coefficients. Combined comparisons of all harmonics to recent experimental data from both RHIC and LHC will potentially allow to determine the desired details of the initial state and the medium properties of the quark-gluon plasma produced in heavy-ion collisions.
\end{abstract}


\section{Introduction}
The observed large elliptic flow at RHIC and recently at the LHC is one of the most striking observations in heavy-ion collision experiments.
This asymmetry of particle production in the transverse plane of the collision is interpreted as the hydrodynamic response to the initial geometry.
The applicability of hydrodynamics demands a short mean free path with respect to the system size. Therefore it is concluded that the created quark-gluon plasma is strongly coupled and behaves like a nearly perfect fluid.
Recently interest has cascaded to all higher harmonics, including odd ones such as the triangular flow coefficient $v_3$, which are non-zero in single events. A lot of progress has been made in extracting medium properties from hydrodynamic calculations, with the largest uncertainty emerging from the limited knowledge of the initial conditions.

\section{Theoretical framework}
The current standard for the viscous hydrodynamic description of relativistic heavy-ion collisions has been established in \cite{Baier:2007ix}, 
where all terms up to second order in gradients for a conformal fluid have been derived. Additional terms for non-conformal fluids with non-zero bulk
viscosity have been derived in \cite{Betz:2009zz}.

In the ideal case, the evolution of the system created in relativistic heavy-ion collisions
is described by the following 5 conservation equations
\begin{eqnarray}
& \partial_\mu T_{\rm id}^{\mu\nu} = 0\,, ~~\partial_\mu J_B^\mu = 0\,, \label{conservationEqns}
\end{eqnarray}
where $T_{\rm id}^{\mu\nu}$ is the energy-momentum tensor and $J_B^\mu$ is the net baryon
current. These are usually re-expressed
using the time-like flow four-vector $u^\mu$ as
\begin{eqnarray}\label{Tideal}
& T_{\rm id}^{\mu\nu} = (\varepsilon + \mathcal{P})u^\mu u^\nu - \mathcal{P} g^{\mu\nu}\,, ~~ J_{B}^{\mu} = \rho_{B} u^\mu \,,
\end{eqnarray}
where $\varepsilon$ is the energy density, $\mathcal{P}$ is the pressure, $\rho_B$
is the baryon density and 
$g^{\mu\nu} = \hbox{diag}(1, -1, -1, -1)$ is the metric tensor.
The equations are then closed by adding the equilibrium 
equation of state
\begin{eqnarray}\label{eos}
\mathcal{P} = \mathcal{P}(\varepsilon, \rho_B)
\end{eqnarray}
as a local constraint on the variables.

In the first-order, or Navier-Stokes formalism for viscous hydrodynamics,
the stress-energy tensor is decomposed into
$
T_{\rm 1st}^{\mu\nu} = T^{\mu\nu}_{\rm id} + S^{\mu\nu}\,,
$
where
$T_{\rm id}^{\mu\nu}$ is given in Eq.\,(\ref{Tideal}) and the viscous part of the stress energy tensor is
given by 
\be
S^{\mu\nu} = \eta 
\left(
\nabla^\mu u^\nu + \nabla^\nu u^\mu - 
{2\over 3}\Delta^{\mu\nu}\nabla_\alpha u^\alpha
\right)\,,
\ee
where $\Delta^{\mu\nu} = g^{\mu\nu} - u^\mu u^\nu$ is the local 3-metric 
and $\nabla^\mu = \Delta^{\mu\nu}\partial_\nu$ is the local spatial derivative.
Note that $S^{\mu\nu}$ is transverse with respect to the flow velocity since
$\Delta^{\mu\nu}u_\nu = 0$ and $u^\nu u_\nu = 1$. 
Hence, $u^\mu$ is also an
eigenvector of the whole stress-energy tensor with the same eigenvalue
$\epsilon$. $\eta$ is the shear viscosity of the medium.

The Navier-Stokes form is conceptually simple but introduces unphysical 
super-luminal signals 
that lead to numerical instabilities. 

The second-order Israel-Steward formalism \cite{Israel:1976tn,Stewart:1977,Israel:1979wp}
avoids this super-luminal propagation, as does the more recent approach in 
\cite{Muronga:2001zk}. 
In the Israel-Stewart formalism for a conformal fluid,
derived in \cite{Baier:2007ix}, the stress-energy tensor is decomposed as
$
{\cal T}^{\mu\nu} = T^{\mu\nu}_{\rm id} + \pi^{\mu\nu}\,.
$
The evolution equations are
\begin{eqnarray}
~~~~~~~~~~~\partial_\mu {\cal T}^{\mu\nu} &=& 0\,,\\
\Delta^{\mu}_{\alpha}\Delta^{\nu}_{\beta}
{u^\sigma\partial_\sigma} \pi^{\alpha\beta}
&=&
-{1\over \tau_\pi}
\left( \pi^{\mu\nu} - S^{\mu\nu} \right) - {4\over 3}\pi^{\mu\nu}(\partial_\alpha u^\alpha)\,,
\end{eqnarray}
where we neglected vorticity and terms that turn out to be numerically irrelevant. For the role of vorticity in heavy-ion collisions when including
fluctuations see \cite{Floerchinger:2011ww}.

Simulations of bulk dynamics in heavy-ion collisions using this formalism have been performed in 2+1 dimensions in \cite{Romatschke:2007jx,Romatschke:2007mq,Luzum:2008cw,Heinz:2005bw,Song:2007fn,Song:2008si}, and within the equivalent
\"Ottinger-Grmela \cite{Grmela:1997zz} formalism in \cite{Dusling:2007gi}. Recently, 3+1 dimensional viscous calculations have also become available \cite{Schenke:2010rr,Schenke:2011tv}.

Having established the standard theoretical framework, 
in the following I will discuss recent developments in the field of relativistic viscous hydrodynamics and its application to heavy-ion collisions.


\section{Equation of state}
Typically, the equation of state (\ref{eos}) used in hydrodynamic simulations of heavy-ion collisions is determined from lattice QCD calculations combined with a hadron gas model. A recent parametrization and a detailed comparison of different lattice equations of state has been performed in \cite{Huovinen:2009yb}.
The different equations of state \cite{Laine:2006cp,Chojnacki:2007jc,Chojnacki:2007rq,Bazavov:2009zn,Song:2008si}, which have different 
results for the trace anomaly and the speed of sound, lead to different evolution of the momentum anisotropy when used in an ideal hydrodynamic evolution.
However, the difference in the final spectra and elliptic flow results turns out to be negligible. This leads to the conclusion that one cannot determine details about the equation of state from the comparison of hydrodynamic calculations to experimental data. On the other hand, using one of the latest lattice equations of state will do a good enough job when extracting medium properties such as transport coefficients from hadronic observables.
Thermal photon production, however, is potentially more sensitive to the equation of state, because photons are produced throughout the evolution and hence probe the dynamics of the system more directly \cite{Gale:2011}. Another relevant aspect is the inclusion of chemical freeze-out in the equation of state (and the freeze-out procedure) to reproduce the correct final particle ratios \cite{Hirano:2002ds,Kolb:2002ve,Shen:2010uy}. 

\section{Bulk viscosity and limits of second order viscous hydrodynamics}
QCD is a non-conformal theory and contains a finite bulk viscosity. In \cite{Denicol:2009am} the effect of bulk viscosity on elliptic flow 
and in \cite{Monnai:2009ad,Song:2009rh} the combined effect of shear and bulk viscosity has been studied. Bulk viscosity is expected to peak (possibly along with its relaxation time) around the critical temperature $T_c$, where the system may develop large correlation lengths \cite{Paech:2006st,Kharzeev:2007wb,Meyer:2009jp,Buchel:2009hv}.
In \cite{Song:2009rh}, within the range of applicability of second order viscous hydrodynamics,
corrections from bulk viscosity are found to be small compared to those from shear viscosity if the bulk relaxation time peaks around $T_c$.
This means that if bulk and shear viscous correction do not become so large as to render the hydrodynamic expansion invalid for the relevant part of a heavy-ion collision, the extraction of the shear viscosity should be possible to reasonable accuracy when neglecting the effect of bulk viscosity. 

Apart from viscous corrections to the thermal equilibrium distribution functions becoming large, bulk and shear viscous corrections can lead to a negative longitudinal pressure, and hence the breakup of the system into droplets \cite{Torrieri:2008ip,Rajagopal:2009yw,Bhatt:2011kr}.
Second order viscous hydrodynamics can hint at when such cavitation happens but is not suited to describe the process as it happens outside the range of its applicability.

\section{Progress beyond second order viscous hydrodynamics}
The problem with the viscous correction to the stress energy tensor becoming larger than the equilibrium part, leading to negative pressure,
arises especially in the early stage of the evolution, when the local momentum distribution is not yet equilibrated but highly anisotropic due to rapid longitudinal expansion. Recent progress has been made in describing this early time evolution and late time hydrodynamics within the same framework by performing the hydrodynamic expansion around an anisotropic distribution \cite{Martinez:2010sc,Martinez:2010sd,Florkowski:2010cf,Ryblewski:2011aq}. 
This results in new equations of motion, including one for the degree of anisotropy of the distribution function. 
The procedure can reproduce both the limits of free streaming and ideal hydrodynamics and results in second order viscous hydrodynamics 
when expanding around a small anisotropy parameter, which so far has been shown in the one dimensional case \cite{Martinez:2010sc}.

Another logical step is to expand to third order in gradients as done in \cite{El:2009vj}. Numerical differences to second order Israel Stewart theory
are completely negligible for $\eta/s=0.05$ and become significant for $\eta/s\gtrsim 0.2$. It has also been pointed out that the hydrodynamic equations depend on the details of their derivation. While Israel and Stewart used the second moment of the Boltzmann equation to derive hydrodynamic equations for the dissipative currents, in \cite{Denicol:2010xn} the definition of the latter was used directly. This leads to equations of motion of the same form but with different coefficients. 
Microscopic transport calculations \cite{Xu:2004mz,Xu:2007ns} show very good agreement with the new equations of motion up to $\eta/s\sim 3$, while the Israel-Stewart equations show large differences for $\eta/s\gtrsim 0.2$. This demonstrates that the details of the derivation are relevant in particular when hydrodynamics is being matched to kinetic theory at late times.

\section{Temperature dependent $\eta/s$}
What is extracted from experimental data by comparison with viscous hydrodynamic calculations using constant $\eta/s$ is at best an effective, or average, $\langle \eta/s \rangle$. In reality $\eta/s$ should depend on the local temperature of the medium, dropping from large values at high temperatures to a minimum at $T_c$, and rising with decreasing temperature in the hadronic phase \cite{Nakamura:2004sy,NoronhaHostler:2008ju,Denicol:2010tr,Niemi:2011ix}.

It has been shown that when including such modeled temperature dependence in calculations at RHIC energies, the details of $\eta/s(T)$ in the quark-gluon plasma phase have little influence on the final elliptic flow result, while hadronic $\eta/s(T)$ modifies $v_2$ strongly \cite{Niemi:2011ix}. At the highest LHC energies the conclusion is the opposite: weak dependence on the hadronic $\eta/s$ but strong dependence on $\eta/s(T)$ in the QGP phase. Interestingly, at RHIC energies, a significant dependence has been found on the minimum value of $\eta/s(T)$ around $T_c$ \cite{Niemi:2011}. This might indicate that we are determining such a  minimal value, rather than $\langle \eta/s \rangle$, when using a constant $\eta/s$.

In addition, it was pointed out in \cite{Shen:2011eg} that there is a strong dependence on the initial value of $\pi^{\mu\nu}$ when starting with a large $\eta/s$. So, particularly when including a temperature dependent $\eta/s$, it is essential to gain a better understanding of the pre-equilibrium stage in heavy-ion collisions to determine the initial conditions for viscous hydrodynamics.

\section{Viscous corrections to particle distribution functions}
When translating the dissipative $T^{\mu\nu}$ to particles in the Cooper-Frye formalism \cite{Cooper:1974mv}, corrections to the distribution function $\delta f$ have to be taken into account:
\begin{equation}
 T^{\mu\nu}_{\rm hydro} = \sum_{n=1}^N d_n \int \frac{d^3p}{(2\pi)^3}\frac{p^\mu p^\nu}{E_n} (f_{0n}+\delta f_n) 
\end{equation}
for an $N$ component system with $d_n$ the degeneracy of species $n$.
 The ansatz
\begin{equation}
  \delta f_n = \frac{C_n}{2T^3}f_0 (1\pm f_0) \hat{p}^\alpha \hat{p}^\beta \chi(p) \frac{\pi_{\alpha\beta}}{\eta}\,,
\end{equation}
where $\hat{p}^\alpha$ is a unit vector in the $\alpha$ direction, leaves some freedom and the usual procedure is to assume that all coefficients $C_n$ are equal, even though they should depend on the individual particle species' interaction rate (see \cite{Molnar:2011}), and use $\chi(p)=p^2$, which is derived within a relaxation time plus Boltzmann approximation:
\begin{equation}
  \delta f_n = f_{0n} (1\pm f_{0n}) p^\alpha p^\beta \pi_{\alpha\beta} \frac{1}{2 (\epsilon+\mathcal{P}) T^2} ~ \forall ~  n\,.
\end{equation}
It has been shown in \cite{Dusling:2009df} that $\chi(p)\propto p^\alpha$, where $\alpha$ can take on values from 1 to 2, which is the case for example for a system with radiative and elastic energy loss that has $\chi(p)\propto p^{1.38}$. The exact form of the correction has a large effect on the $p_T$ differential elliptic flow for $p_T\gtrsim 1\,{\rm GeV}$. It should be noted, however, that the analysis of experimental data in \cite{Lacey:2010fe} indicates $\chi(p)\propto p^2$.
A more general problem is that corrections $\delta f$ can become large compared to $f_0$. For hadrons this problem can be reduced when using a hadronic afterburner and switching at intermediate temperatures of $\sim 160\,{\rm MeV}$, but for photons that are produced throughout the whole evolution this becomes a serious concern \cite{Gale:2011}.

\section{Hadronic afterburner}
The use of hybrid models coupling early hydrodynamic evolution to a microscopic hadronic cascade in 
the later stage has a long history \cite{Bass:2000ib,Teaney:2000cw,Hirano:2005xf,Nonaka:2006yn,Nonaka:2010zz,Hirano:2010jg,Hirano:2010je}.
Recently 2+1 dimensional viscous hydrodynamics has been coupled to UrQMD \cite{Song:2010aq,Song:2010mg,Song:2011hk,Song:2011qa,Heinz:2011}.
When using a chemically frozen equation of state, transverse momentum spectra of produced particles show very little dependence on the temperature
$T_{\rm sw}$ at which the switching between hydro and UrQMD is performed. However, $v_2$ does show a dependence unless an increasing $\eta/s(T)$ is introduced in the hadronic stage of the hydro simulation to emulate the effect of UrQMD's larger dissipation \cite{Song:2010aq}. This $\eta/s$ rises to $\sim 0.4$, for which the applicability of second order viscous hydrodynamics becomes questionable. Also, it turns out that the dynamics in UrQMD cannot be described
by viscous hydrodynamics with $\eta/s(T)$ due to different relaxation times \cite{Song:2010aq}, underlining the need for the more realistic microscopic rescattering.

\section{Initial conditions and event-by-event hydrodynamics}
Fluctuating initial conditions for hydrodynamic simulations of heavy-ion collisions
have been argued to be very important for the exact determination of collective flow observables and to describe 
features of multi-particle correlation measurements in heavy-ion collisions
\cite{Andrade:2006yh,Adare:2008cqb,Abelev:2008nda,Alver:2008gk,Alver:2009id,Abelev:2009qa,Miller:2003kd,Broniowski:2007ft,Andrade:2008xh,Hirano:2009bd,
Takahashi:2009na,Andrade:2009em,Alver:2010gr,Werner:2010aa,Holopainen:2010gz,Alver:2010dn,Petersen:2010cw,Schenke:2010rr,Schenke:2011tv,Qiu:2011iv}.
Real event-by-event hydrodynamic simulations have been performed and show modifications to spectra and flow from ``single-shot'' 
hydrodynamics with averaged initial conditions \cite{Holopainen:2010gz,Schenke:2010rr,Schenke:2011tv,Qiu:2011iv,Jeon:2011}.
An important advantage of event-by-event hydrodynamic calculations is the possibility to study higher flow harmonics such as $v_3$,
which are entirely due to fluctuations in the initial conditions. Different $v_n$ depend differently on $\eta/s$ and the details of the initial condition, 
which is determined by the dynamics and fluctuations of partons in the incoming nuclear wave functions.
This observation can be used to determine these long sought after
details of the initial state and medium properties in heavy-ion collisions by performing a systematic analysis
of all harmonics $v_n$, up to e.g. $n=6$ as a function of $\eta/s$ and the initial state properties and compare to experimental data.
First predictions of $v_3$ \cite{Schenke:2010rr} agree extremely well with experimental data from RHIC \cite{Adare:2011tg}.
Furthermore, it has now been shown that at low $p_T$ (and $|\Delta \eta|>1$ (ALICE), $|\Delta \eta|>2$ (ATLAS)), the main features of dihadron correlations in the angular difference $\Delta \phi$ between the hadron momenta can be described by flow, i.e. the sum of $v_1$ to $v_6$ only \cite{Jia:2011,ALICE:2011vk}. 
The double-peak structure on the away-side is hence described mostly by (triangular) flow as predicted in \cite{Takahashi:2009na,Alver:2010gr}.

\vspace{-0.2cm}
\section{Summary of LHC predictions,  conclusions and outlook}
Calculations using ideal and viscous hydrodynamics at LHC energies hint at little or no increase of $\langle \eta/s\rangle$ 
\cite{Luzum:2009sb,Luzum:2010ag,Hirano:2010jg,Hirano:2010je,Schenke:2011tv,Song:2011qa,Roy:2011xt}. 
However, to be consistent a temperature dependent $\eta/s$ should be employed to extend RHIC calculations to LHC (see e.g. \cite{Song:2011qa}). 
Also, essentially all models underestimate $v_2(p_T)$
for $p_T\lesssim 0.8\,{\rm GeV}$, which might be explained by contributions from non-thermalized particles \cite{Bozek:2011wa}. The issue is however not settled yet.
Nevertheless, the overall early success of hydrodynamics 
in describing the main features of flow of charged hadrons at the LHC indicates that the QGP at LHC is also a nearly perfect fluid.
Higher harmonics are also being computed for the LHC and comparison of experimental data to first predictions from event-by-event simulations \cite{Schenke:2011tv} show that
hydrodynamic results are in the right ball park. As mentioned earlier, a systematic analysis will be needed to potentially determine $\eta/s(T)$ and the details of the initial conditions at LHC with the use of all $v_n$.

Better understanding of viscous corrections to the distribution functions and dependence on the model for coupling to a hadronic afterburner,
detailed studies of 3+1 dimensional hydrodynamics with viscosity, and above all a better understanding of
the pre-equilibrium stage and its transition to hydrodynamics are the next important steps in developing a reliable viscous hydrodynamic description of the bulk dynamics in heavy-ion collisions.

\section*{Acknowledgments}
I thank Charles Gale, Sangyong Jeon, Derek Teaney, and Raju Venugopalan for helpful comments on the manuscript and discussions.
This work was supported in part by the US Department of Energy under DOE Contract No.DE-AC02-98CH10886, and 
by a Lab Directed Research and De\-velopment Grant from Brookhaven Science Associates.

\section*{References}
\bibliography{hydro}

\end{document}